\documentclass[prd,showpacs,letterpaper,twocolumn]{revtex4}%
\usepackage{graphicx}
\usepackage{bm}
\usepackage{epsf}
\usepackage{rotating}
\usepackage{epsfig,graphics,rotate,color}
\usepackage{wrapfig}
\usepackage{amssymb}
\usepackage{amsmath}
\usepackage{amsfonts}
\usepackage{array,hhline,dcolumn}%
\providecommand{\U}[1]{\protect\rule{.1in}{.1in}}
\begin{document}
\title{A Decisive Disappearance Search at High-\boldmath$\Delta m^2$ with Monoenergetic Muon Neutrinos}
\author{S. Axani$^1$, G. Collin$^{1}$, J.M. Conrad$^{1}$,  M.H. Shaevitz$^{2}$,
  J. Spitz$^1$, T. Wongjirad$^1$}
\affiliation{$^{1}$ Massachusetts Institute of Technology, Cambridge, MA 02139, USA}
\affiliation{$^{2}$ Columbia University, New York, NY 10027, USA}

\begin{abstract}

``KPipe" is a proposed experiment which will study muon neutrino disappearance for a sensitive test of the $\Delta m^2\sim1~\mathrm{eV}^2$ anomalies, possibly indicative of one or more sterile neutrinos. The experiment is to be located at the J-PARC Materials and Life Science Facility's spallation neutron source, which represents the world's most intense source of charged kaon decay-at-rest monoenergetic (236~MeV) muon neutrinos.    The detector vessel, designed to measure the charged current interactions of these neutrinos, will be 3~m in diameter and 120~m long, extending radially at a distance of 32~m to 152~m from the source. This design allows a sensitive search for $\nu_\mu$ disappearance associated with currently favored light sterile neutrino models and features the ability to reconstruct the neutrino oscillation wave within a single, extended detector. The required detector design, technology, and costs are modest. The KPipe measurements will be robust since they depend on a known energy neutrino source with low expected backgrounds. Further, since the measurements rely only on the measured rate of detected events as a function of distance, with no required knowledge of the initial flux and neutrino interaction cross section, the results will be largely free of systematic errors.  The experimental sensitivity to oscillations, based on a shape-only analysis of the $L/E$ distribution, will extend an order of magnitude beyond present experimental limits in the relevant high-$\Delta m^2$ parameter space.
\end{abstract}

\maketitle

\section{Introduction}

A number of experimental anomalies consistent with neutrino oscillations at a characteristic mass splitting around 1~eV$^2$ hint at the possibility of an additional neutrino. These anomalies fall into two categories:  muon-to-electron flavor appearance, as observed by the LSND~\cite{LSND} and MiniBooNE~\cite{MBnu, MBnubar} experiments, and electron flavor disappearance, as observed by reactor~\cite{reactor1,reactor2} and source~\cite{Gallium,Giuntixsec,SAGE3,GALLEX3} experiments.  A favored beyond-Standard-Model explanation for these anomalies invokes an additional number of $N$ sterile neutrinos participating in oscillations beyond the three active flavors~\cite{sorel, sbl, kopp, giunti}.  These  ``3+$N$ models''  are able to simultaneously describe the existing anomalous observations and those measurements which do not claim a signal in the relevant parameter space~\cite{KARMEN,MB_SB,ConradShaevitz,MBnudisapp,MBNumi,NOMAD,CCFR84,CDHS,atmos}, although there is tension between both neutrino and antineutrino measurements and appearance and disappearance measurements.     
% An important outcome of these global studies is that, given the existence of a new neutrino oscillation frequency or frequencies, muon-flavor disappearance is predicted to arise close to the edge of what has been probed in $\nu_\mu$ disappearance to date.  % similar sentence repeated a paragraph later.

Muon neutrinos, for example, must disappear if the observed anomalies are due to oscillations involving a light sterile neutrino. The lack of observed $\nu_\mu$ disappearance is a major source of tension in the global fits. In order to understand the importance of $\nu_\mu$ disappearance measurements, consider a 3+1 sterile neutrino model, with the probability for $\nu_\mu \rightarrow \nu_e$ appearance given by:
% This fact, alongside the lack of actual observed $\nu_\mu$ disappearance in this parameter space, motivates the construction of a fast, decisive, and low cost $\nu_\mu$ disappearance experiment.   % again repeated
\begin{equation}
P(\nu_{\mu}\rightarrow\nu_{e})\simeq  4|U_{\mu 4}|^2 |U_{e 4}|^2\sin^2(1.27\Delta m^2_{41}L/E)~.\label{app}
\end{equation}
 The probability for $\nu_e$ and $\nu_\mu$  disappearance are, respectively:
\begin{equation}
P(\nu_{e}\rightarrow\nu_{e})\simeq
1-4(1-|U_{e 4}|^2)|U_{e 4}|^2\sin^2(1.27\Delta m^2_{41}L/E)~ \label{disappeq2}
\end{equation}
and
\begin{equation}
P(\nu_{\mu}\rightarrow\nu_{\mu})\simeq
1-4(1-|U_{\mu 4}|^2)|U_{\mu 4}|^2\sin^2(1.27\Delta m^2_{41}L/E)~. \label{disappeq1}
\end{equation}
In these equations,  the elements of the mixing matrix, $U$, set the amplitude of oscillation, while $\Delta m^2_{41}$ establishes the oscillation wavelength. Within this 3+1 model, a global fit to the world's data, including all anomalies and null results, will simultaneously constrain $U_{e 4}$, $U_{\mu 4}$, and $\Delta m^2_{41}$. The range of values that $U_{\mu 4}$ can take on, and therefore the oscillation parameters that govern $\nu_\mu$ disappearance, can thus be restricted.    The present global fit for $\nu_\mu$ disappearance places the allowed region just outside of current bounds. This motivates the construction of a fast, low cost, and decisive $\nu_\mu$ disappearance experiment that can confirm or disallow various models for sterile neutrinos, as well as inform a range of future proposed experiments \cite{whitepaper, isodar, prospect, nulat, sox,jparc, fnal_sbn, coherent,oscsns,isodar_kamland_juno}.

In what follows we describe such an experiment, called ``KPipe", that can perform a search for $\nu_\mu$ disappearance that extends well beyond current limits while still being low cost. KPipe will employ a long, liquid scintillator-based detector that is oriented radially with respect to an intense source of isotropic monoenergetic 236~MeV $\nu_\mu$s coming from the decay-at-rest of positively charged kaons ($K^+\rightarrow\mu^+\nu_{\mu}$; BR=63.55$\pm$0.11\%~\cite{pdg}).  As the only relevant monoenergetic neutrino that can interact via the charged current interaction, a kaon decay-at-rest (KDAR) $\nu_{\mu}$ source represents a unique and important tool for precision oscillation, cross section, and nuclear physics measurements~\cite{kdar1,kdar2}.  Since the energy of these neutrinos is known, indications of $\nu_{\mu}$ disappearance may be seen along the length of the KPipe detector as oscillating deviations from the expected 1/$R^{2}$ dependence in the rate of $\nu_{\mu}$ charged-current (CC) interactions. A measurement of such a deviation over a large range of $L/E$ would not only be a clear indication for the existence of at least one light sterile neutrino, but also begin to disambiguate among different sterile neutrino models.  % I want to lead this paragraph with introducing KPipe because it better connects to the introduction of an experimental opportunity

%The result, shown in Fig.~\ref{fig:KPipe90msens}, is a
%90\%~CL sensitivity limit (gray region)
%which is an order of magnitude greater than existing $\nu_\mu$
%disappearance limits~\cite{numudis} (dotted black).    

\section{The KDAR source and KPipe Detector Design}

%%(Maybe this section should broadly describe the way the experiment is setup and the way it works.  Later we describe our simulation with all of the exact choices for cuts and numbers.  This way its more clear that our simulation is not an exact description of the experiment whose final details we can't possibly have decided on now, but rather serves as a proof of principle that we can achieve the goals we want.)

\begin{figure}
\centering
\begin{minipage}{1\linewidth}
  \includegraphics[width=\linewidth]{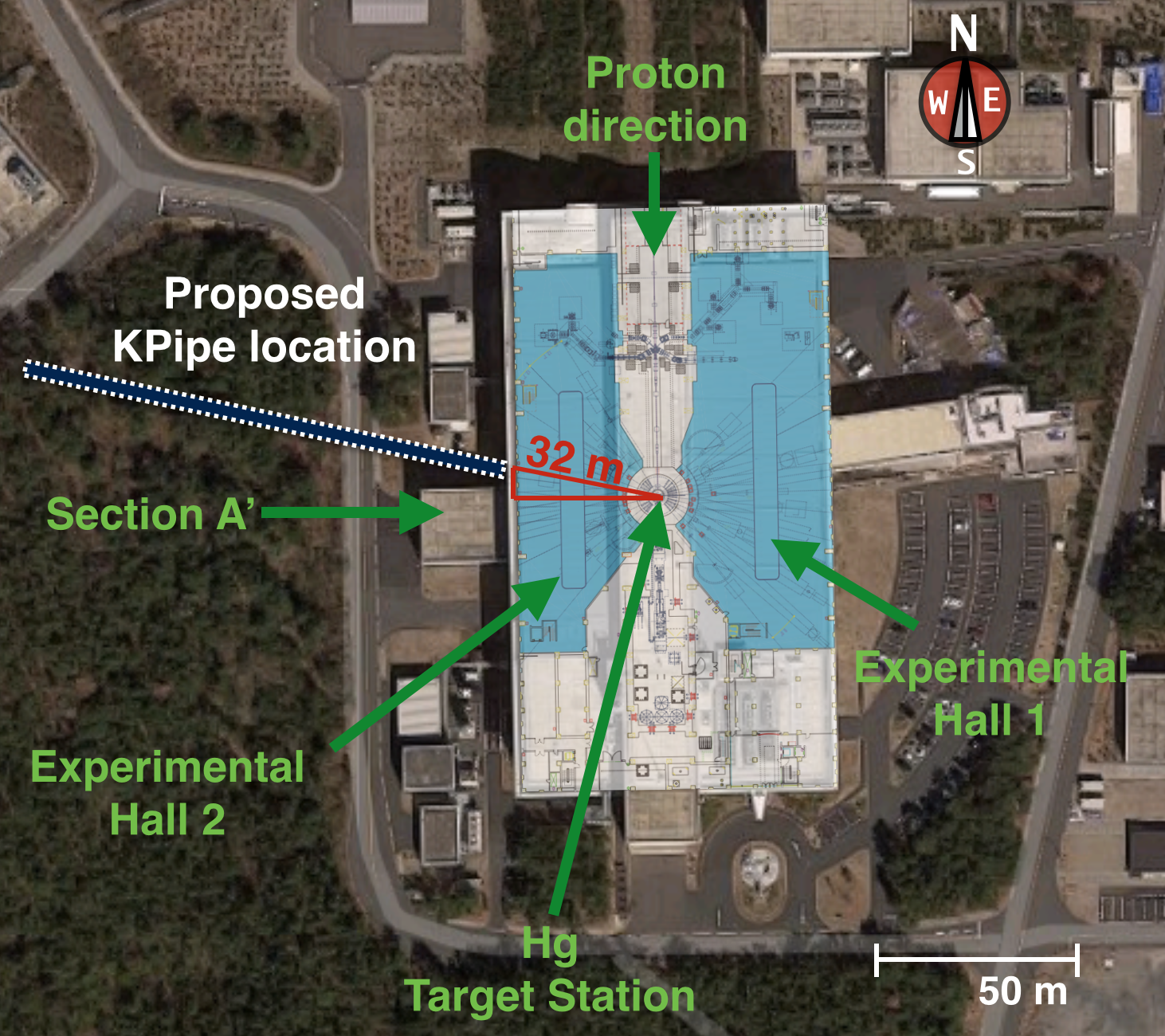}
\end{minipage}
\caption{An aerial view from Google Maps (2015) of the Materials and Life Science Facility layout with a superimposed schematic drawing~\cite{jparc} of the first floor, including the target station. The proposed KPipe location (shown with a dotted contour) is 32~m from the target station and 102$^\circ$ with respect to the incident proton beam direction. The detector extends radially outward from the target station.}
\label{fig:mlf}
\end{figure}

The Materials and Life Science Experimental Facility (MLF) at the Japan Proton Accelerator Research Complex (J-PARC) in Tokai, Japan houses a spallation neutron source used for basic research on materials and life science, as well as research and development in industrial engineering.  It is also an intense, yet completely unutilized, source of neutrinos that emits the world's most intense flux of KDAR monoenergetic (236 MeV) $\nu_\mu$s.  Neutron beams along with neutrinos are generated when a mercury target is hit by a pulsed, high intensity proton beam from the J-PARC rapid-cycling synchrotron (RCS)~\cite{jparc}. The RCS delivers a 3~GeV, 25~Hz pulsed proton beam, which arrives in two 80~ns buckets spaced 540~ns apart. The facility provides users 500~kW of protons-on-target (POT) but has demonstrated its eventual steady-state goal of 1~MW, albeit for short times~\cite{mlf_news}. Along with neutrons, the proton-on-target interaction provides an intense source of light mesons, including kaons, which usually come to rest in the high-A target and surrounding shielding. 

\begin{figure}
\centering
\begin{minipage}{1\linewidth}
  \includegraphics[width=\linewidth]{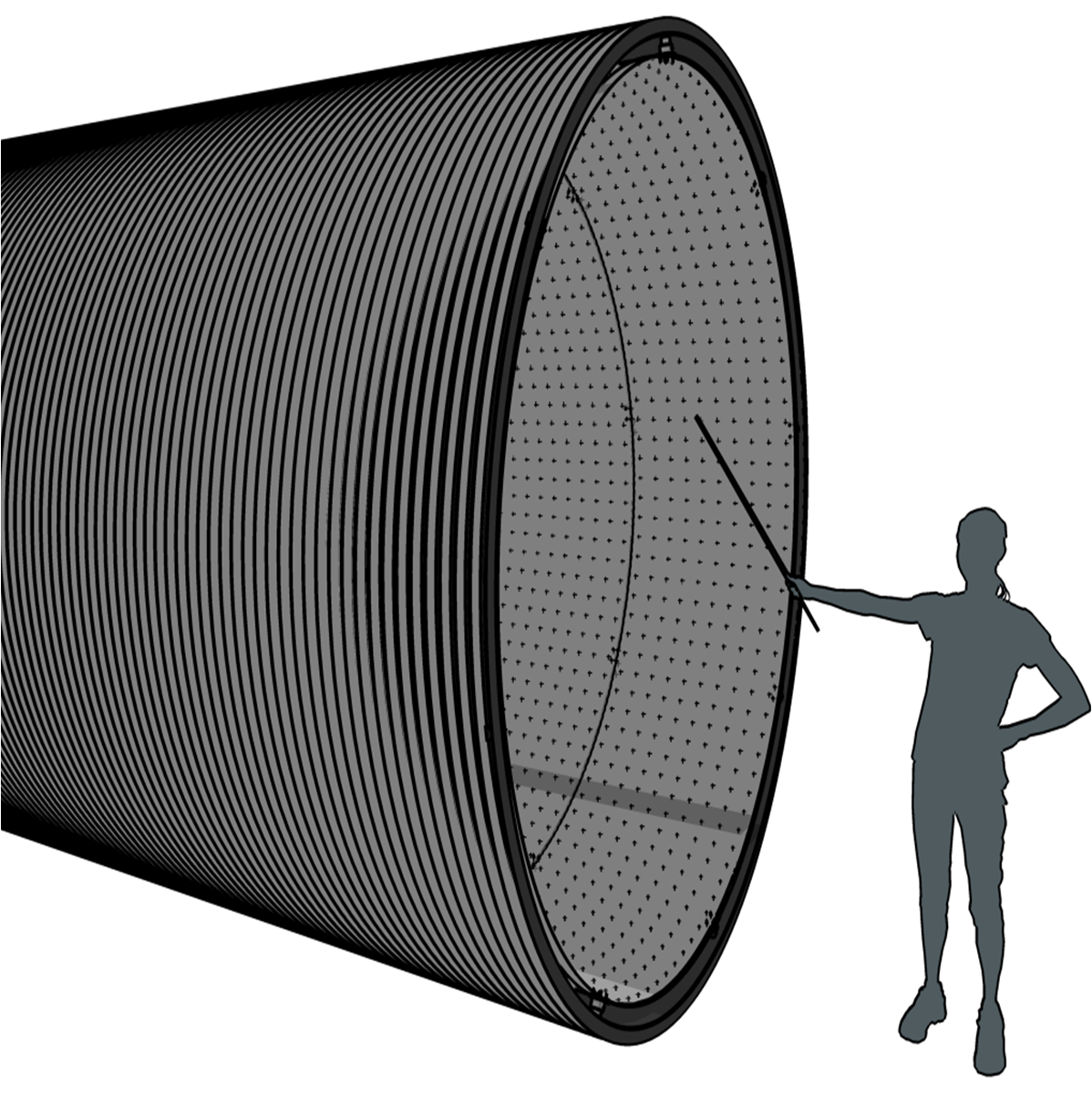}
\end{minipage}
\caption{The KPipe detector design, featuring a 3~m inner diameter high density polyethylene (HDPE) vessel filled with liquid scintillator. Silicon photomultipliers (SiPMs) are seen mounted on the interior panels in hoops spaced by 10~cm in the longitudinal direction. The cosmic ray veto is a 10~cm space between the panels and the outer HDPE wall.}
\label{fig:pipe}
\end{figure}

%Charged current interactions of $\nu_{\mu}$s on carbon nuclei ($\nu_\mu\mathrm{^{12}C} \rightarrow \mu^- X$) provide a signal channel to measure the flux of neutrinos coming from the MLF source.  In a liquid scintillator detector, this interaction produces a visible muon and $X$, where $X$ is some combination of an excited nucleus, de-excitation photons, and one or more ejected nucleons after final state interactions. The muon can then decay to produce a detectable electron.  

KPipe will search for muon-flavor disappearance with CC interactions of 236 MeV $\nu_{\mu}$s on carbon nuclei ($\nu_\mu\mathrm{^{12}C} \rightarrow \mu^- X$) in liquid scintillator. This interaction produces a visible muon and $X$, where $X$ is some combination of an excited nucleus, de-excitation photons, and one or more ejected nucleons after final state interactions. The goal of the KPipe detector design is to efficiently identify these 236~MeV $\nu_\mu$ CC events, broadly characterized by two separated flashes of light in time coming from the prompt $\mu^- X$ followed by the muon's decay electron. %We note that the difference between a before-final-state-interactions pure ``quasi-elastic" (QE) event ($\nu_\mu n \rightarrow \mu^-  p$) and an ``absorption" event ($\nu_\mu  \mathrm{^{12}C} \rightarrow \mu^- X$) is blurred in this transition energy region. While this does complicate the theoretical treatment of neutrino scattering at this energy, uncertainties in the cross section are not important for the disappearance measurement as only the relative interaction rate along the detector is relevant.  

% what does the sentence on QE versus absorption want to achieve?
%While this does complicate the theoretical treatment of neutrino scattering at this energy, the contribution from other interaction modes, such as resonant excitation and deep inelastic scattering, is negligible. 
%Signs of muon-flavor disappearance are searched for through CC interactions of 236 MeV $\nu_{\mu}$s on carbon nuclei ($\nu_\mu\mathrm{^{12}C} \rightarrow \mu^- X$). We note that the difference between a before-final-state-interactions pure ``quasi-elastic" (QE) event ($\nu_\mu n \rightarrow \mu^-  p$) and an ``absorption" event ($\nu_\mu  \mathrm{^{12}C} \rightarrow \mu^- X$) is blurred in this transition energy region. While this does complicate the theoretical treatment of neutrino scattering at this energy, the contribution from other interaction modes, such as resonant excitation and deep inelastic scattering, is negligible. 

The KPipe design calls for a relatively low cost, 3~m inner diameter (ID) steel-reinforced, high-density polyethylene (HDPE) pipe that is filled with liquid scintillator. As shown in Fig.~\ref{fig:mlf}, the pipe is positioned so that it extends radially outward from the target station. The upstream location maximizes the sensitivity to oscillations by being the shortest possible distance from the source, given spatial constraints.  We have found that a long detector (120~m, 684~tons) is most suitable for optimizing sensitivity to oscillations across a wide range of the most pertinent parameter space, in consideration of current global fit results, the neutrino energy, 1/$R^2$, and estimated cost.

The interior of the pipe contains a cylinder of highly reflective panels, which optically separate the active volume from the cosmic ray (CR) veto. Hoops of inward-facing silicon photomultipliers (SiPMs) are mounted on the interior of the panels. There are 100~equally-spaced SiPMs per hoop, and each hoop is separated longitudinally by 10~cm (see Fig.~\ref{fig:pipe}). The space surrounding the inner target region on the other side of the panels is the 10~cm-thick veto region. The surfaces of the veto region are painted white, or lined with a Tyvek$^{\textregistered}$-like material, for high reflectivity.  Along the innermost side of the veto region are 120 hoops of outward-facing SiPMs that each run along the circumference of the pipe.  The hoops have 100~SiPMs each and are positioned at 1 m spacing along the inside of the veto region.  The 10~cm spaces at the ends of the pipe are also instrumented.  Each veto end cap is instrumented with 100 SiPMs that all face axially outward and are spaced equally apart on a circle with 1~m radius.

SiPMs are employed in both the target and veto regions because of their compact size and reduced cost when purchased in bulk.  Currently available SiPMs typically have a quantum efficiency around 30\%.  In order to further reduce cost, we plan on multiplexing the SiPM channels.  For the target region, each channel of readout electronics monitors 25 out of the 100~total SiPMs on a hoop.  For the veto region, one channel monitors one side or end cap hoop.  The active area of a SiPM can range from 1~mm$^{2}$ to about 6~mm$^{2}$.  Assuming 6~mm$^{2}$ SiPMs, with 1200~hoops containing 100~SiPMs each, the target region will have a photocathode-coverage of only $\sim0.4$\%.  Despite this low coverage, simulations of the experiment described in the next section indicate that there are an adequate number of SiPMs to achieve the goals of the experiment.  %Future work will investigate the use of alternative light collection systems, such as an array of PMTs, which could increase photocathode-coverage for a similar total cost.

%For SiPM quantities needed to cover the KPipe detector, the active photocathode-coverage per unit cost is comparable to that of photomultiplier tubes (PMTs). However, SiPMs have an advantage in that they are much more compact than PMTs. One reason for this advantage is the reduced amount of dead volume in the detector. Though even for PMTs, this will be small relative to the large target volume. More importantly, the SiPMs' smaller size also means that, keeping costs fixed, the photocathode-coverage can be better spread out along the surface of a detector using SiPMs. This is important due to the KPipe detector's long length. However, even with 1200~hoops containing 100~SiPMs each, the target region will still have very low photocathode-coverage, only around 0.4\%. Given this low coverage, the KPipe detector therefore relies on a liquid scintillator detector medium to produce a large amount of light per unit of energy deposited.  

The KPipe detector succeeds despite the sparse amount of instrumentation in the inner region because of its use of liquid scintillator as the detector medium. The low photocathode coverage is overcome by the large amount of light produced by the scintillator per unit of energy deposited.  Scintillators under consideration for KPipe include those based on mineral oil and linear alkylbenzene (LAB).  One example of a currently-deployed mineral oil-based scintillator is the one used by the NO$\nu$A experiment~\cite{nova}.    This scintillator is a mixture of 95\%-by-mass mineral oil with 5\% pseudocumene (1,2,4-trimethylbenzene) along with trace amounts of PPO (2,5-diphenyloxazole) and bis-MSB (1,4-Bis(2-methylstyryl)benzene) wavelength shifters~\cite{nova_scint}. The UV photons emitted by the pseudocumene excite the PPO, which, as the primary scintillant, re-emits in the range of 340-380~nm. These photons are then absorbed by the bis-MSB and reemitted in the 390-440~nm range. Along with developing their scintillator, the NO$\nu$A experiment has also established the methods to manufacture large quantities of it at a relatively low cost.  Other examples of mineral oil-based scintillators are those offered by Saint-Gobain.  For reference, the light yield of these scintillators range from 28\% to 66\% of anthracene or $\sim$4500 to $\sim$11400~photons/MeV~\cite{saintgobain}.  Besides mineral oil, another option is to use a LAB-based liquid scintillator, similar to that being used by the  SNO+ experiment~\cite{snoplab}.  This liquid scintillator consists of the LAB as solvent with PPO acting as the wavelength shifter.  The advantage of a LAB-based liquid scintillator over those based on mineral oil is that it has a comparable light yield to the brighter Saint-Gobain scintillators~\cite{LABly} while also being less toxic. In order to be conservative, we assume in simulations of the KPipe detector (discussed in the next section) a light yield consistent with the dimmest mineral oil based liquid scintillator from Saint-Gobain (4500~photons/MeV). The liquid scintillator that is eventually employed for KPipe will be some optimization between light yield, cost, and safety.  %T: I still think the 4500 photon number is out of place here.

\section{Simulation of the Experimental Setup}

In order to study the performance capabilities of KPipe, we have created simulations of both the neutrino source and the detector.  The source simulations, using both Geant4~\cite{geant4} and MARS15~\cite{mars}, model 3~GeV kinetic energy protons hitting the mercury target. The resulting particles are propagated, and the kinematics of all the neutrinos produced are recorded. A semi-realistic geometry is employed with Geant4 for the target and surrounding material, although the majority (86\%) of 236~MeV $\nu_\mu$ are found to originate within the mercury target. About 75\% of the $K^+$ are found to DAR within 25~cm of the upstream end of the mercury target and the ratio of $\nu_\mu$ from $K^+$ DAR to $\nu_\mu$ from $K^+$ decay-in-flight over 4$\pi$ is $\sim$13:1. The $K^+$ production rate varies depending on which simulation software is used. The Geant4 model calculates the 236~MeV $\nu_\mu$ yield to be 0.0038~$\nu_\mu$ per proton on target (POT), whereas the MARS15 model predicts 0.0072~$\nu_\mu$/POT. Later, when calculating the sensitivity of the experiment in Section~\ref{sec:sens}, we will quote a sensitivity which relies on the MARS15 model for kaon production, as it has been more extensively tuned to data than Geant4~\cite{marsvalid}.  

\begin{figure*}
 	\center
  	\begin{minipage}{.45\linewidth}
  	\includegraphics[width=\linewidth]{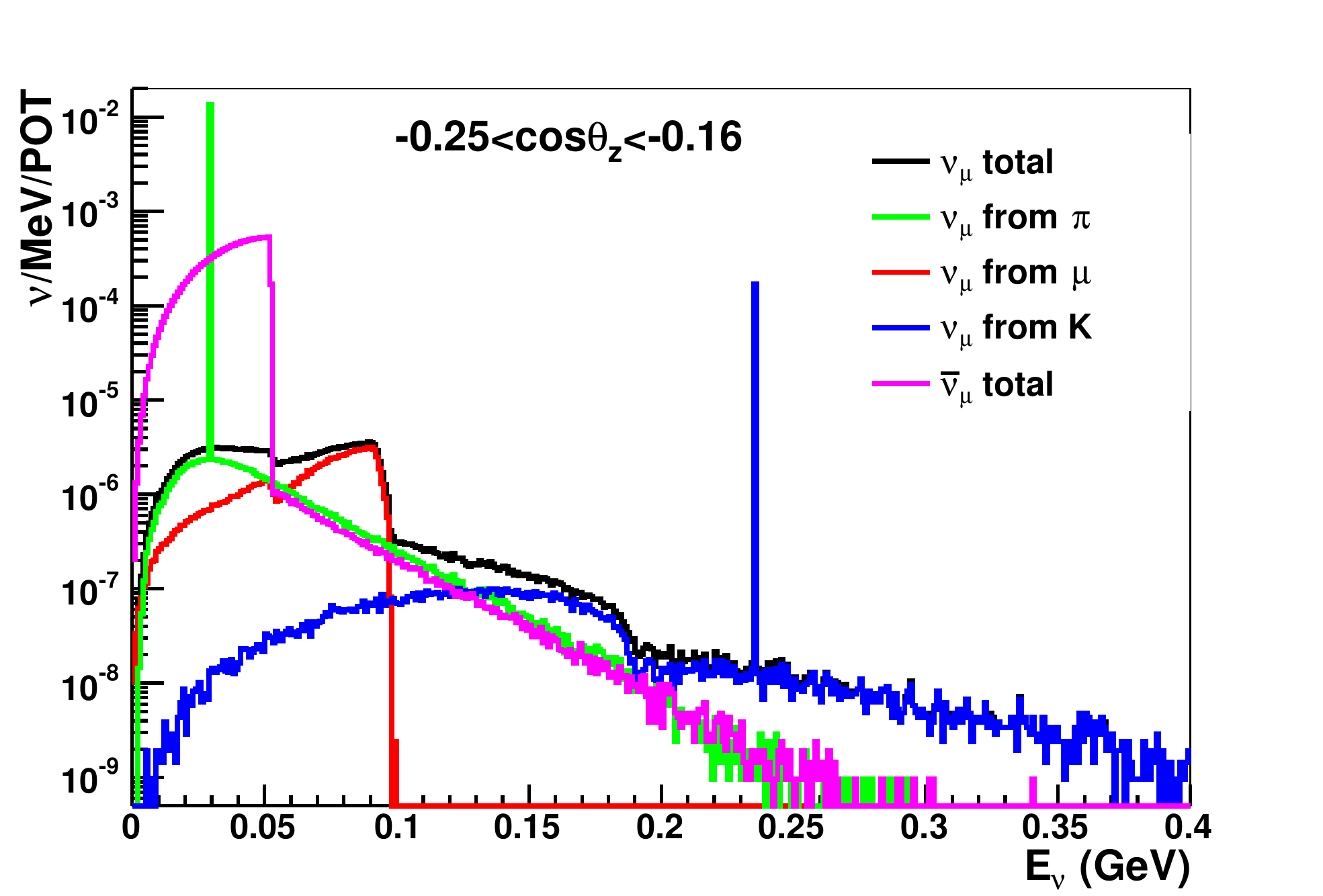}
\end{minipage}
\hspace{.05\linewidth}
\begin{minipage}{.45\linewidth}
 \includegraphics[width=\linewidth]{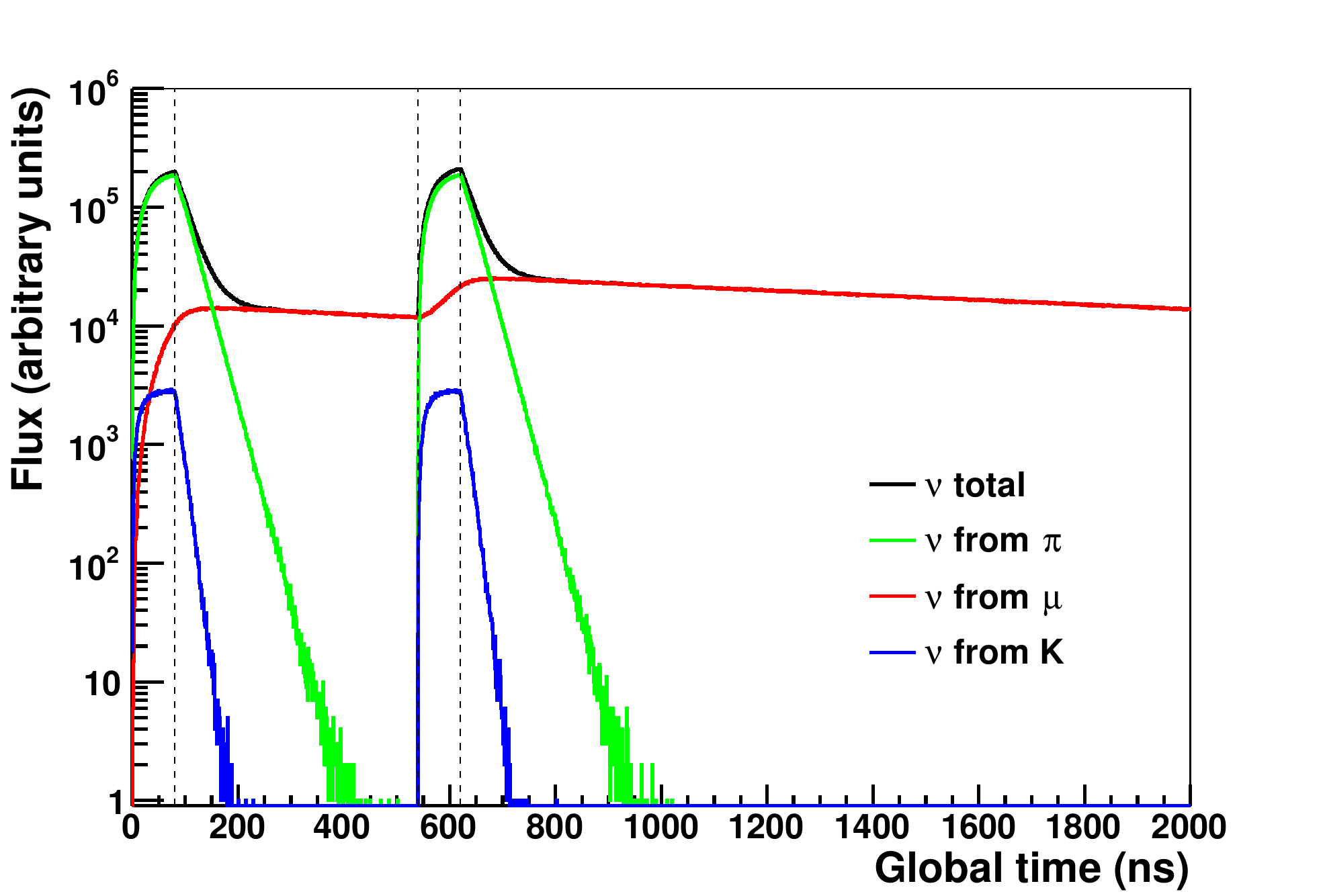}
	\end{minipage}
  	\caption{Left: The muon neutrino and antineutrino flux with $-0.25<\cos \theta_{z}<-0.16$, representative of the full detector length, where $\theta_{z}$ is the neutrino angle with respect to the proton direction (+$z$). Right: The neutrino creation time relative to the two beam pulses (dotted lines).  This distribution includes neutrinos emitted over all solid angles and energies.}
  	\label{fig:flux}
\end{figure*}

The $\nu_\mu$ flux is propagated to the KPipe detector whose closest end to the source is 32~m away. The $\nu_\mu$ flux for $-0.25<\cos \theta_{z}<-0.16$, where $\theta_{z}$ is the neutrino angle with respect to the proton direction (+$z$), representative of the full detector length, is shown in Fig.~\ref{fig:flux} (left). The time distribution of all neutrinos coming from the source is shown in Fig.~\ref{fig:flux} (right). The two 80~ns wide proton pulses can be seen in the figure, while the blue histogram shows the neutrinos coming from kaon decay. 

\begin{figure}
  \centering
  \begin{minipage}{\linewidth}
  \includegraphics[width=1.0\textwidth]{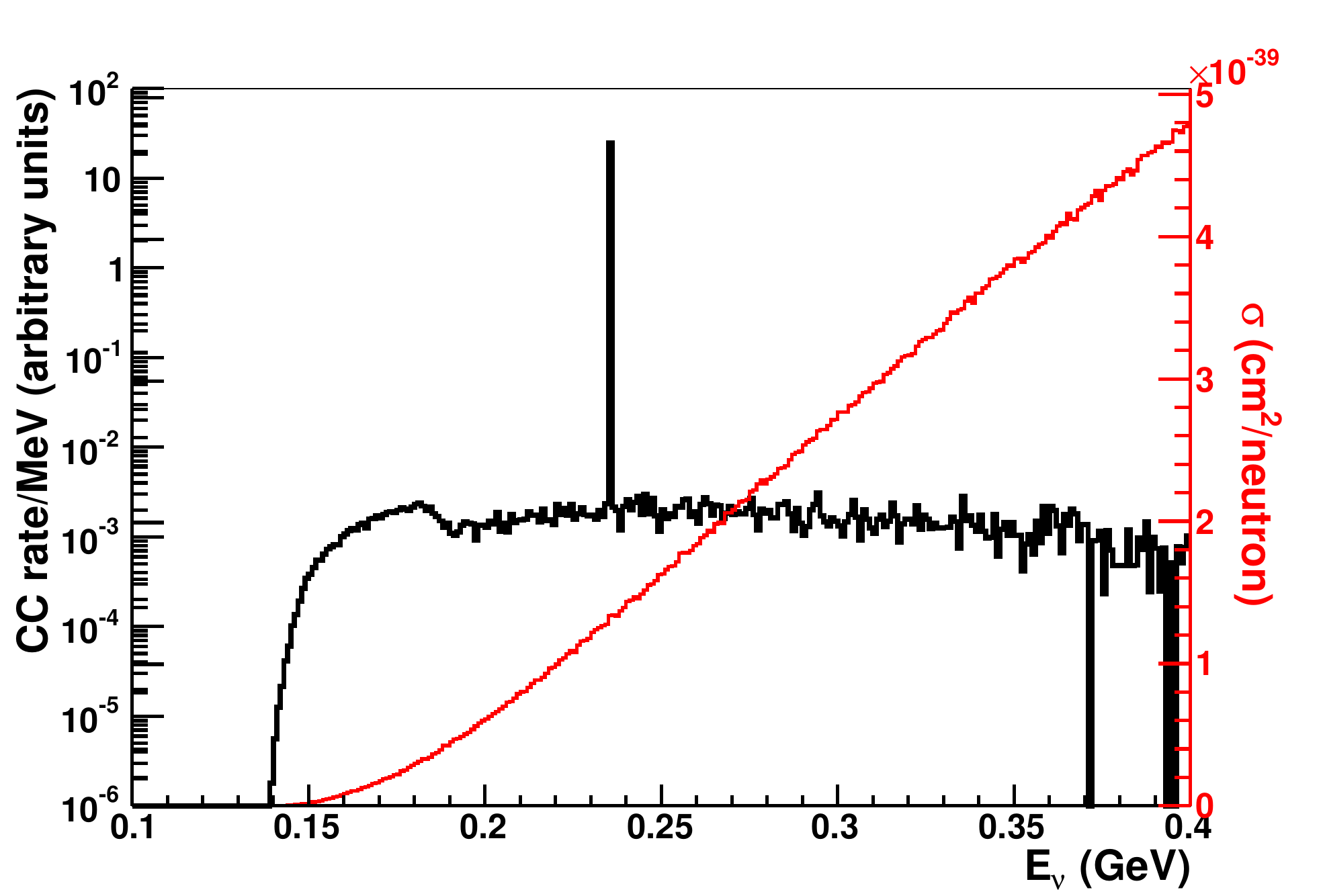}
  \caption{The $\nu_{\mu}$ charged current event rate, for neutrinos with $-0.25<\cos \theta_{z}<-0.16$, along with the employed $\nu_\mu$ CC cross section. The monoenergetic 236 MeV neutrino signal is clearly visible above the ``background" non-monoenergetic events, mainly coming from kaon decay-in-flight.}
    \label{fig:rate}
  \end{minipage}
\end{figure}

The interactions of neutrinos with the detector and surrounding materials are modeled with the NuWro event generator~\cite{NuWro}, and the $\nu_\mu$ CC cross section and expected rate can be seen in Fig.~\ref{fig:rate}. Notably, the signal (KDAR) to background (non-KDAR) ratio is 66:1 integrated over all energies. In other words, if a neutrino-induced muon is observed, there is a 98.5\% chance that it came from a 236~MeV $\nu_\mu$ CC interaction. Given 5000 hours/year of J-PARC 1~MW operation (3.75 $\times 10^{22}$~POT/year), consistent with Ref.~\cite{bkg}, we expect $1.02\times10^5$ KDAR $\nu_\mu$ CC events/year in the 684~ton active volume.

\begin{figure}
  \centering
  \begin{minipage}{\linewidth}
  \includegraphics[width=1.0\textwidth]{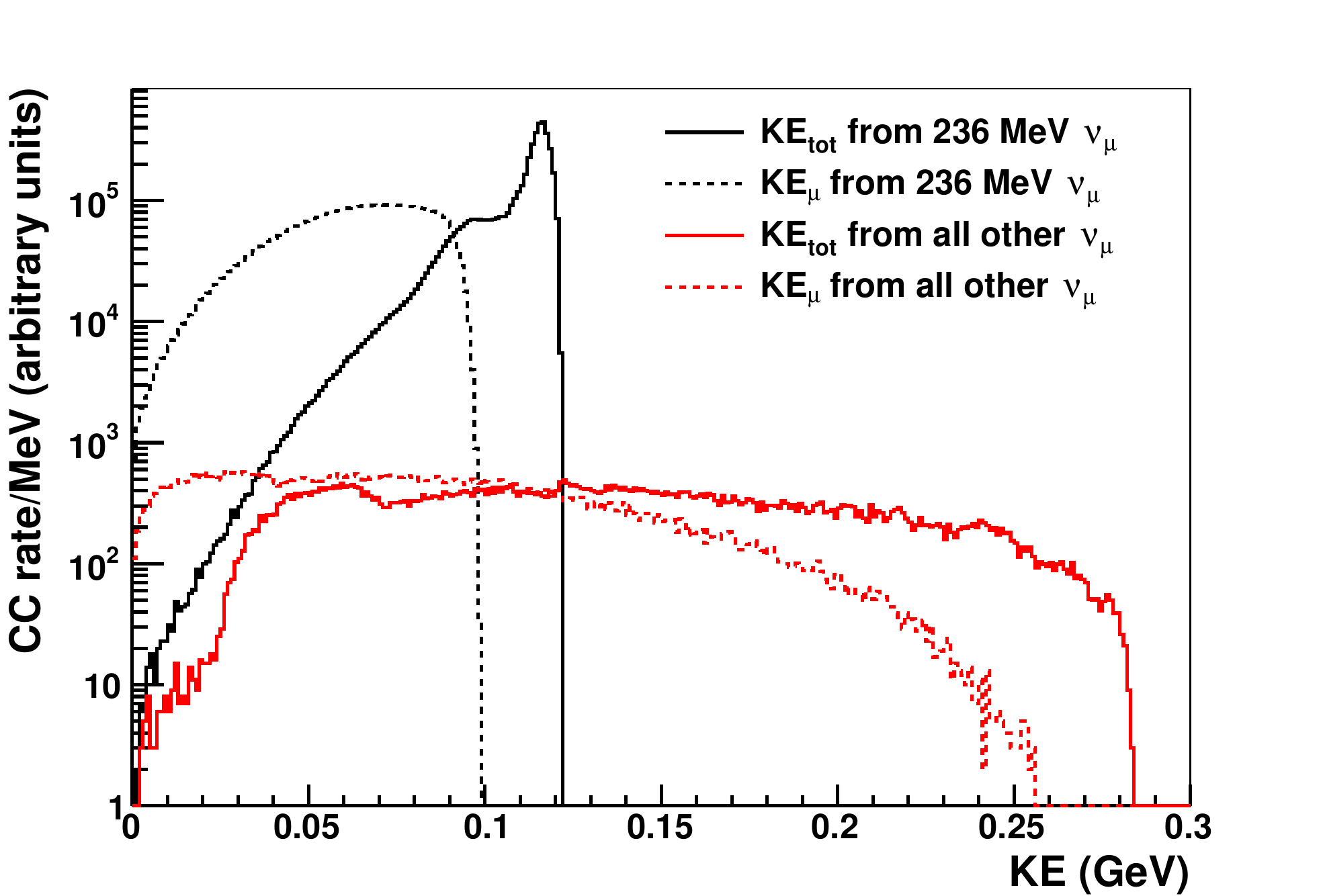}
  \caption{The muon and total kinetic energy ($\mathrm{KE}_\mathrm{tot}=\mathrm{KE}_\mu+\sum{\mathrm{KE}_\mathrm{p}}$) for the signal 236 MeV $\nu_\mu$ charged current events compared to all other $\nu_\mu$. Only neutrinos with $-0.25<\cos \theta_{z}<-0.16$ are considered. The ratio of integrated signal (black) to background (red) is 66:1.}
    \label{fig:nuwro_muon_kinematics}
  \end{minipage}
\end{figure}

For each generated 236~MeV $\nu_\mu$ CC interaction on carbon, NuWro provides the momentum of the outgoing muon and any final state nucleons (typically a single proton). Fig.~\ref{fig:nuwro_muon_kinematics} shows the kinetic energies of the resulting KDAR signal muons along with the non-KDAR muons. The $\nu_\mu$ CC cross section on carbon at 236~MeV according to NuWro and employed for the event rate estimate here is $1.3\times10^{-39}~\mathrm{cm}^2/\mathrm{neutron}$. This is consistent with the Random Phase Approximation (RPA) model's~\cite{martini1,martini2,martiniprivate} cross section prediction of $(1.3+0.2) \times10^{-39}~\mathrm{cm}^2/\mathrm{neutron}$ (RPA QE+npnh). While NuWro is the only generator we use to produce simulated events, we did compare the kinematic distributions given by NuWro to that provided by GENIE~\cite{genie} and the Martini \textit{et al.} RPA model~\cite{martiniprivate}, which includes multi-nucleon effects.  We find that the difference in the muon kinematic predictions among the models is not large enough to significantly change the detector simulation and oscillation sensitivity results. 

Particle propagation through the detector is modeled using the Geant4-based simulation package RAT~\cite{rat}. The detector geometry input into the simulation is as described in the previous section.  The detector is assumed to be on the surface and is surrounded by air only.  Neutrino events are distributed over a 5~m~x~5~m~x~140~m box that fully contains the 120~m long, 3~m diameter cylindrical detector.  The distribution of events in the box is weighted to take into account the 1/R$^{2}$ dependence of the flux along with the density of the various materials in the simulation.   The small divergence in the neutrino direction is also considered.  The RAT package includes a model for scintillator physics that derives from models previously employed by other liquid scintillator experiments such as KamLAND.  The processes that are considered include scintillation, absorption, and reemission.  All three have wavelength dependence. The reflectivity of surfaces in the detector is simulated using the models built into Geant4. 

In addition to the simulation of KDAR neutrino interactions with the detector and surrounding material, we simulate the propagation of CR throughout the volume.  We use the simulation package CRY~\cite{cry} to study the CR particle flux, which generates showers consisting of some combination of one or more muons, pions, electrons, photons, neutrons, or protons.  The dark rate of SiPMs is also included in the simulation of the SiPM response.  We use a dark rate of 1.6~MHz for each of the 130,200 4~mm x 4~mm SiPMs (0.4\% photo-coverage) along with a total quantum efficiency of 30\%. The dark rate comes from the specification for SenSL series C SiPMs which have an advertised dark rate of $<100,000$ Hz/mm$^2$~\cite{sensl_seriesc}.

\section{Isolating and Reconstructing $\nu_\mu$ Events from the KDAR Source}
\label{sec:recon}

Signal events from the KDAR neutrino source are identified by the observation of two sequential pulses of light. The first pulse comes from the muon and vertex energy deposition. The next signal is from the Michel electron produced by the decay of the muon ($\nu_\mu\mathrm{^{12}C} \rightarrow \mu^- X, \mu^- \rightarrow e^- \nu_\mu \overline{\nu}_e$). We apply a pulse finding algorithm to identify both light signals from the SiPMs.  The algorithm uses a rolling 20~ns window over which the number of hits in the SiPMs are summed and the expected dark hit contribution in the window is subtracted.  The first pulse is found when the hit sum with subtraction is above a given threshold, specifically one that is four times larger than the standard deviation of the expected number of dark hits.  After the first pulse is found, the algorithm searches for the Michel signal using the same method, except that the threshold is raised to account for both the expected dark noise and the contribution of SiPM hits from the first pulse.  This expected hit contribution is dictated by the decay time of the scintillator.  After isolating coincident signals, the position along the detector of both the primary interaction and Michel signal is determined by the photoelectron-weighted position of the hits seen by the SiPMs.  Using the position of the prompt pulse, we find that the vertex position resolution of the interaction is 80~cm.  The current proposed readout is likely unable to reconstruct more detailed information about the event such as the muon angle, although this information is not necessary for KPipe's primary measurement. 

\begin{figure}
  \centering
  \begin{minipage}{\linewidth}
  \includegraphics[width=1.0\textwidth]{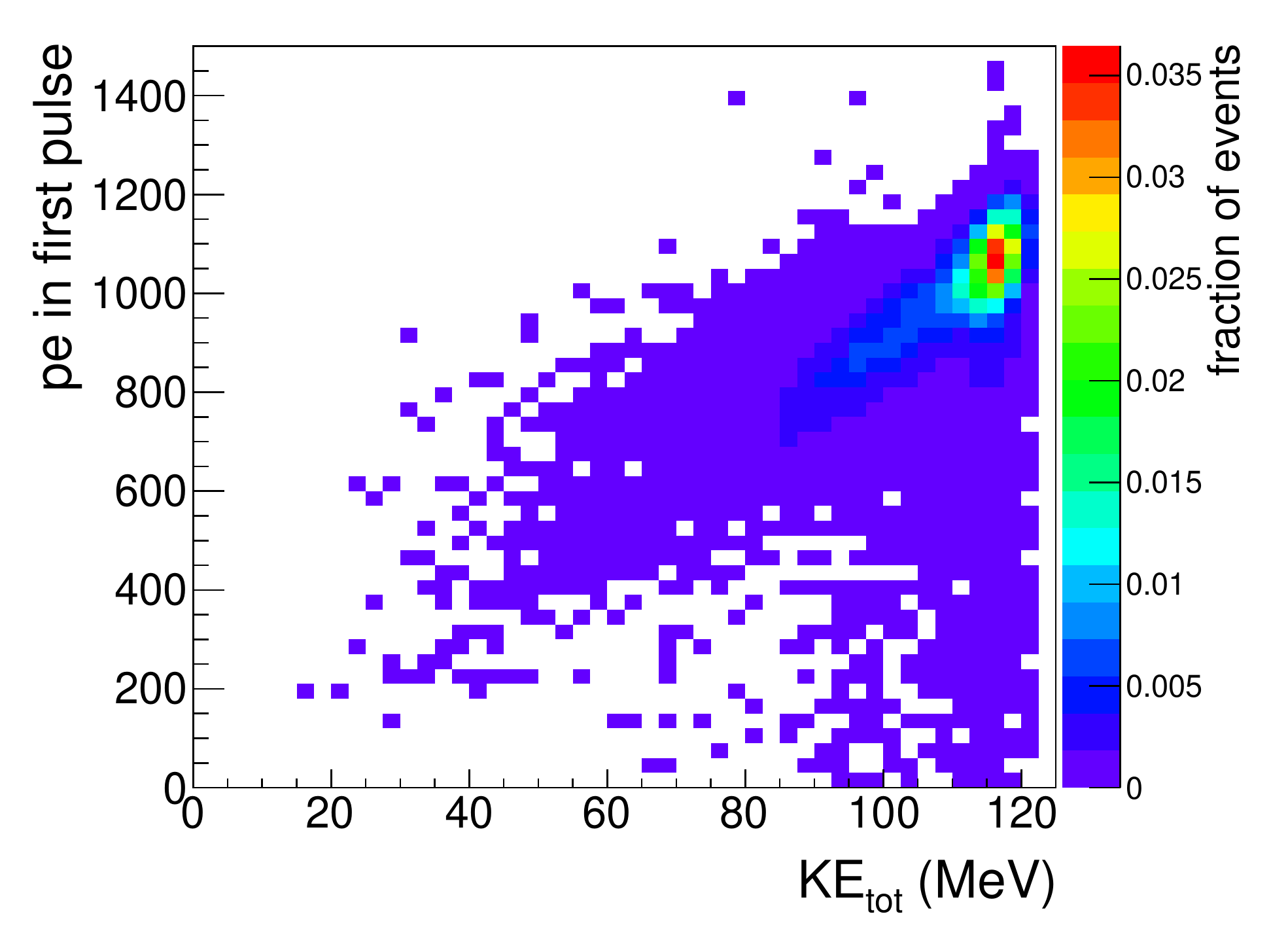}
  \caption{The number of photoelectrons in a 236 MeV $\nu_\mu$ CC event's first pulse versus the total kinetic energy ($\mathrm{KE}_\mathrm{tot}=\mathrm{KE}_\mu+\sum{\mathrm{KE}_\mathrm{p}}$).}
    \label{fig:energy_scale}
  \end{minipage}
\end{figure}

Fig.~\ref{fig:energy_scale} shows the number of photoelectrons ($pe$) in the first pulse as a function of total kinetic energy, $\mathrm{KE}_{\mathrm{tot}}$, defined as the total kinetic energy of the muon and any final state protons ($\mathrm{KE}_\mathrm{tot}=\mathrm{KE}_\mu+\sum{\mathrm{KE}_\mathrm{p}}$). The figure shows simulated data from 236~MeV KDAR $\nu_\mu$ CC interactions.  The first pulse usually contains over 800~$pe$, indicating that, despite the low photocathode coverage, the amount of observed light for the signal events is high enough for efficient reconstruction.  Further, the figure shows that KE$_{\textrm{tot}}$ correlates well with the number of $pe$ seen. Using the peak of this distribution, the detector light yield is calculated to be 9.2~$pe$/MeV, which includes effects from quantum efficiency, photocathode-coverage, and absorption. %The neutrino energy is not directly related to the $KE_{\mathrm{tot}}$ but is known to be 236 MeV for KDAR neutrinos.

\subsection{Isolating the Signal} 
\label{sec:signal}

The primary background to the $\nu_{\mu}$ CC signal events comes from stopping cosmogenic muons in the detector. We envision applying the following selection requirements in order to select signal interactions and reject CR backgrounds:
\begin{enumerate}
\item the interaction signal (prompt) occurs within 125~ns windows following each of the two 80~ns beam pulses, 
\item the interaction signal has a reconstructed energy in the range $22<E_{\mathrm{vis}}<142$~MeV ($200<pe<1300$),
\item the Michel signal occurs within 10~$\mu$s of the first pulse,   
\item the Michel signal reconstructed visible energy is $11<E_{\mathrm{vis}}<82$~MeV ($100<pe<750$), 
\item the distance between the interaction signal and the Michel signal is less than 1.5~m, and 
\item the summed pulse height in the ten nearest veto SiPM hoops to the interaction signal is less than four times the dark rate $\sigma$ within a 125~ns window after the start of each 80~ns beam pulse.
\end{enumerate}
Note that for the cuts on visible energy, $E_{\mathrm{vis}}$, the corresponding values in $pe$ are given in parentheses.  These are the values used in the Monte Carlo study of the KDAR signal efficiency and CR background rejection.

The first cut (1) takes advantage of the pulsed proton beam. Accepting events only within a 125~ns window after each 80~ns proton pulse efficiently selects 99.9\% of the KDAR neutrinos 
while removing many of the events coming from other neutrino sources. The small 125~ns
event window also limits the rate of CR ray events even before the other selection cuts are applied.
%According to the simulation, CR particles create at least one detectable flash in either the target region or veto in only 0.65\% of all windows.
According to the simulation, CR particles create at least one detectable flash in either the target region or veto in only 0.87\% of all windows.
%The rate of comic ray events that pass the selection criteria is 20~Hz, estimated using the cosmic ray particle simulation in combination with RAT and Geant4.

\begin{figure}
\centering
\begin{minipage}{1\linewidth}
  \includegraphics[width=\linewidth]{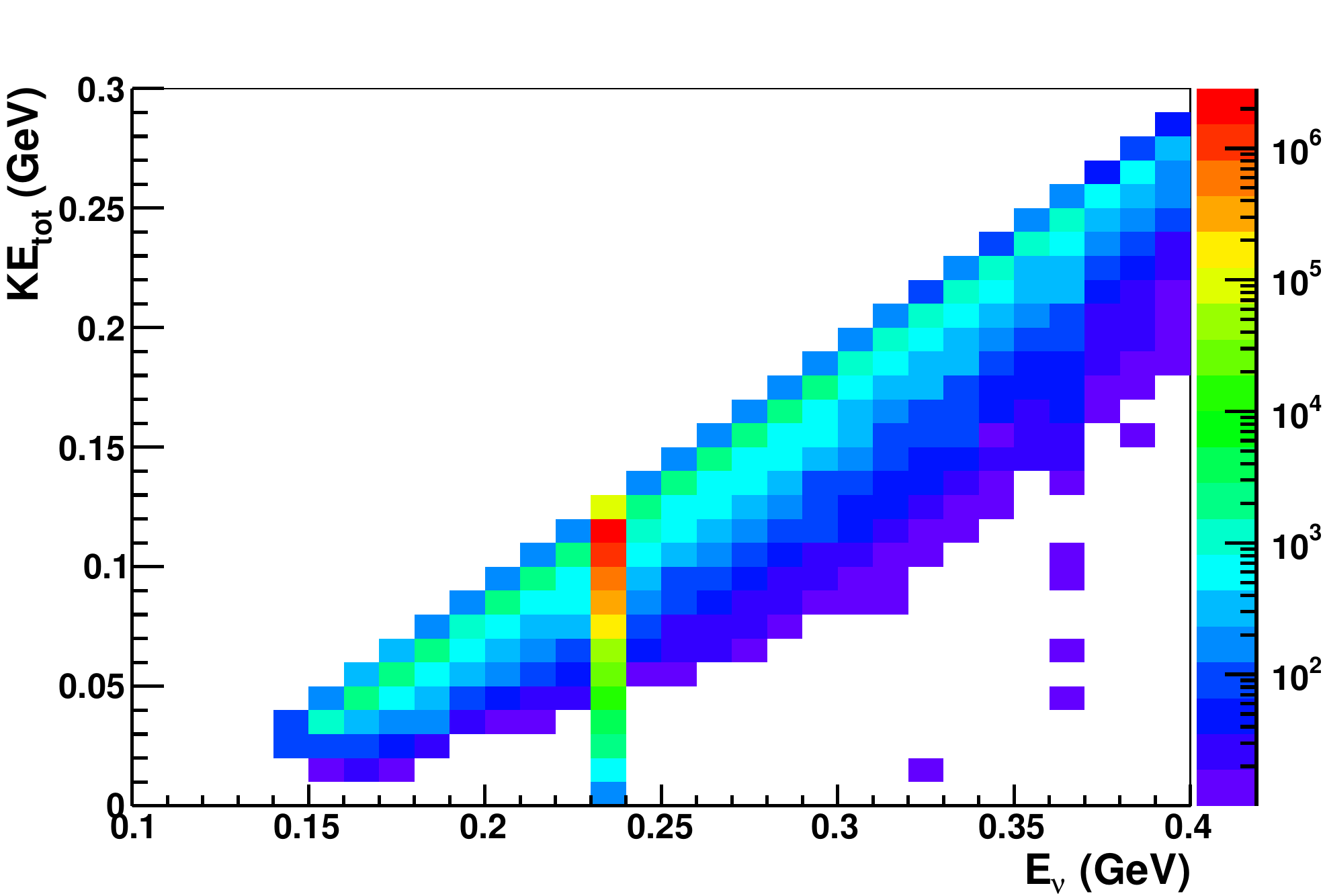}
\end{minipage}
\caption{The total kinetic energy ($\mathrm{KE}_\mathrm{tot}=\mathrm{KE}_\mu+\sum{\mathrm{KE}_\mathrm{p}}$) versus the energy of neutrinos from CC interactions in KPipe. Only neutrinos with $-0.25<\cos \theta_{z}<-0.16$ are considered. The Z-axis units are arbitrary.}
\label{fig:muon_kinematics}
\end{figure}

The second cut (2) utilizes the fact that, because the signal events come from monoenergetic neutrinos, the energy of the outgoing particles falls in a fairly narrow range. Fig.~\ref{fig:muon_kinematics} shows the total kinetic energy of the muons and any final state protons, $\textrm{KE}_{\textrm{tot}}$, as a function of neutrino energy for $\nu_\mu$ CC events in the detector. The upper bound of 142~MeV ensures that the signal neutrino events are preserved with high efficiency, while removing non-KDAR muon neutrinos at higher energies. More importantly, the upper bound removes bright CR events.   Based on the simulation, 72\% of all detectable CR events (i.e. ones that produce one or more detected flashes) are removed by the high energy cut, many of which are through-going muons. Along with kaon decay-in-flight neutrinos, the low energy bound also removes all relevant backgrounds from CR-induced spallation products and is well above the visible energy from radiogenic backgrounds.  With both a high and low energy cut on the first pulse, 87\% of all CR events are removed.

The cuts related to Michel electron timing, energy, and spatial coincidence (cuts 3-5) are chosen to efficiently retain
signal while removing most of the in-time through-going CR muons that traverse the detector, as
well as other backgrounds.  A coincident signal coming from non-stopping muons
can occur due to a CR shower with two or more particles or an associated muon spallation-induced isotope.   The timing, energy, and spatial cuts on the Michel
candidate reduce much of this coincident background.  Applying the above cuts along with the Michel pulse cuts reduces the CR rate to 750~Hz, 
which means that only 0.01\% of all signal windows will contain a CR event.  
At this stage in the cuts, less than two percent of detectable CR events remain.  

The final cut (6) applied removes all events that create a flash of light in the veto.  
The veto is only 10~cm thick and is more sparsely instrumented than the target region.  
However, enough light is produced that the veto is able to reject 99.5\% of all detectable CR events with at least one muon.  We find that lining the walls of the veto with a highly reflective material plays an important role in the veto performance. With all cuts applied, we estimate that the rate of CR events is 27~Hz over the entire active volume.  

In addition to CR backgrounds and non-KDAR muon neutrino events, an additional coincident background can come from beam-induced neutron interactions that produce a
$\Delta^+$ in the detector that subsequently decays into a $\pi^+$.  The latter can then stop and decay to a muon followed by a Michel electron. We assume that this background is negligible for this study. All in-time beam-related backgrounds will be measured before deploying KPipe, and adequate shielding will be installed in order to mitigate them. 

Overall, our studies indicate that the dominant background is from CR shower events that are not removed by the above cuts.   Of the 27~Hz rate that passes, the simulations show that 70\% of the rate is due to stopping muons. The remaining 30\% is due to showers involving photons, electrons, and neutrons.  In the simulation, we do not include any additional passive shielding, for example coming from overburden.  If the detector is buried or shielded, we expect these non-muon backgrounds to be further reduced. The CR background should be distributed uniformly throughout the detector and can be measured precisely using identified out-of-time stopped muons. As a result, only the statistical error from the total number of background events expected to pass the cuts is included in the sensitivity analysis, described later in Section~\ref{sec:sens}.

\subsection{Detection efficiency}  \label{sec:detection_efficiency}

The cuts introduce inefficiency in the signal.  We assume that the neutrino events are distributed evenly in radius and fall as 1/$R^2$ throughout the detector. Signal events near the lateral edge of the target region can exit the detector before the muon can decay. This leads to an acceptance that is a function of radius. Based on an active detector radius of 1.45~m, we find an acceptance of 87\% with respect to KDAR $\nu_\mu$ CC interactions whose true vertex is in the target region.  The selection cuts described above are 89\% efficient according to the simulation.   This includes events where the muon is captured by the nucleus, which occurs in the target region 6\% of the time.  For a subset of these events, there is also an additional 0.75\% dead-time loss due to the rate of CR events in the veto. 
% in what follows i scaled everything by 0.77/0.75 = 1.026666

In summary, the total efficiency for all signal events is 77\%, leading to an expected total KDAR $\nu_\mu$ CC rate of $7.8\times10^4$~events distributed along the pipe's active volume per year of running. This is on average $4.9\times10^{-5}$ KDAR events per proton beam window without oscillations.  This compares with $3.4\times10^{-6}$ CR events per proton beam window. In the most upstream 1~m of the detector, the unoscillated signal to background ratio is about 60:1; in the most downstream 1~m of the detector, the unoscillated signal to background ratio is about 3:1.
% (previous below)
%In summary, the total efficiency for all signal events is 77\%, leading to an expected total KDAR $\nu_\mu$ CC rate of $7.6\times10^4$~events distributed along the pipe's fiducial volume per year of running. This is on average $4.8\times10^{-5}$ KDAR events per proton beam window without oscillations.  This compares with $3.4\times10^{-6}$ CR events per proton beam window. At the end of the pipe nearest the source, the unoscillated signal to background ratio in the number of events in the first 1 m of the detector is about 60.  At the furthest end of the pipe, the unoscillated signal to background ratio in the last 1 m of the detector is about 3.

%\section{Systematic errors}

\section{Sensitivity}
\label{sec:sens}
\begin{figure}
\centering
\begin{minipage}{1\linewidth}
  \includegraphics[width=\linewidth]{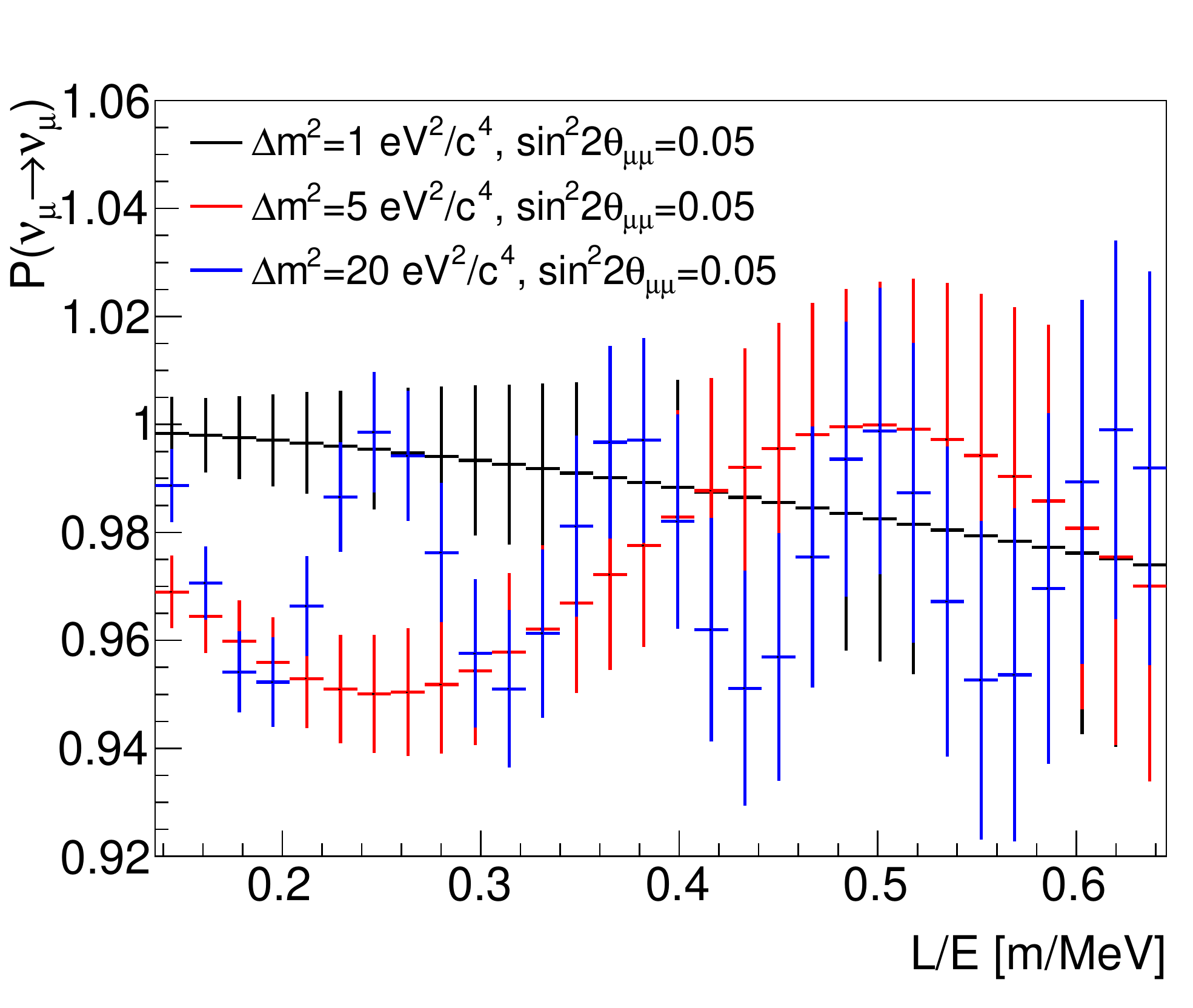}
\end{minipage}
\caption{Three sample oscillation probability measurements as a function of $L/E$ for 3 years of running. The error bars incorporate statistical uncertainties of both the $\nu_{\mu}$ signal and the cosmic ray background.}
\label{fig:oscillations}
\end{figure}

The oscillation probabilities for three different $\Delta m^2$ values ($1, 5, 20~\mathrm{eV}^2$) can be seen in Fig~\ref{fig:oscillations}. The error bars correspond to the statistical uncertainty associated with a 3~year $\nu_{\mu}$ measurement with a CR rate of 27~Hz. This background rate corresponds to 132~CR events that pass our selection cuts for each 1~m slice of the detector.

The sensitivity of the experiment is evaluated using a shape-only $\chi^2$ statistic similar to that described in Ref.~\cite{franke}.  However, we replace the covariance matrix with the Neyman $\chi^2$ convention, since we do not include any correlated systematic uncertainties between each $L/E$ bin. Using Eq.~\ref{disappeq1} for the oscillation probability, the $\chi^2$ value at each pair of oscillation parameters, $\Delta m^2$ and $U_{\mu 4}$, is calculated by comparing the no-oscillation signal ($\mathrm{N}_\mathrm{i}^{\mathrm{\nu,un}}+\mathrm{N}_\mathrm{i}^\mathrm{bkgd}$) to the oscillation signal ($\mathrm{N}_\mathrm{i}^{\mathrm{\nu,osc}}+\mathrm{N}_\mathrm{i}^\mathrm{bkgd}$) in each $L/E$ bin, i. Here, $\mathrm{N}_\mathrm{i}^{\nu\mathrm{,un}}$ and $\mathrm{N}_\mathrm{i}^{\nu\mathrm{,osc}}$ are defined as the number of expected $\nu_\mu$ events in bin i given a no-oscillation prediction and an oscillation prediction, respectively. The number of events in a bin due to background is then added to the $\nu_\mu$ prediction. The $\Delta L$ value used in setting the bin size is 0.5~m. Defining for each  $\textrm{i}^{\mathrm{th}}$ $L/E$ bin the difference between the no-oscillation and oscillation signal, $\mathrm{n}_\mathrm{i}$, where 
\begin{equation} \label{eqn:ni}
\mathrm{n_i} = \left(\mathrm{N}_\mathrm{i}^{\nu\mathrm{,un}}+\mathrm{N}_\mathrm{i}^{\mathrm{bkgd}}\right)-
\left(\xi \mathrm{N}_\mathrm{i}^{\nu\mathrm{,osc}}+\mathrm{N}_\mathrm{i}^{\mathrm{bkgd}}\right),
\end{equation}
the $\chi^2$ is then 
\begin{equation} \label{eqn:chi2}
\chi^2=\sum\limits_{\mathrm{i}}^\mathrm{nbins}\frac{\mathrm{n}_\mathrm{i}^{2}}{\mathrm{N}_\mathrm{i}^{\mathrm{\nu,un}}+\mathrm{N}_\mathrm{i}^{\mathrm{bkgd}}}.
\end{equation}
The normalization constant, $\xi$, in Eq.~\ref{eqn:ni}, is included in order to make the analysis shape-only and is constrained to be
\begin{equation} \label{eqn:xi}
\xi = \frac{\sum\limits_\mathrm{i} \mathrm{N}_\mathrm{i}^{\nu\mathrm{,un}}}{\sum\limits_\mathrm{i} \mathrm{N}_\mathrm{i}^{\nu\mathrm{,osc}}}.
\end{equation}
For the 90$\%$ confidence limit reported, a one degree of freedom, one-sided raster scan threshold of $\chi^2=$1.64 is used. The 5$\sigma$ threshold is $\chi^2=$25.0, considering a one degree of freedom, two-sided raster scan. 

For the subsequent sensitivity plots, the oscillation prediction, $\mathrm{N}_\mathrm{i}^{\nu,\mathrm{osc}}$, has been simplified by the two flavor approximation to the 3+1 neutrino oscillation model (Equation \ref{disappeq1}), where we define $\sin^2(2\theta_{\mu \mu})=4|U_{\mu 4}|^2(1-|U_{\mu 4}|^2)$. 

\begin{table} [t!]
\begin{center}
\begin{tabular}{|c|c|}
\hline
Parameter & Value  \\ \hline
Detector length &  120~m   \\
Active detector radius & 1.45~m   \\
Closest distance to source &  32~m   \\
Liquid scintillator density &  0.863 g/cm$^3$  \\
Active detector mass & 684~tons  \\
Proton rate (1~MW) & 3.75 $\times 10^{22}$ POT/year\\
KDAR $\nu_\mu$ yield (MARS15)  &  0.0072~$\nu_\mu$/POT   \\
$\nu_\mu$ CC $\sigma$ @ 236 MeV (NuWro)  &  $1.3\times10^{-39}~\mathrm{cm}^2/\mathrm{neutron}$   \\
Raw KDAR CC event rate &  $1.02\times10^5$ events/year \\
KDAR signal efficiency &    77\% \\
Vertex resolution  &   80~cm \\
Light yield & 4500~photons/MeV \\
$\nu_{\mu}$ creation point uncertainty &    25~cm\\
Cosmic ray background rate &    27~Hz \\
\hline
\end{tabular} \caption{Summary of the relevant experimental parameters.}\label{table:values}
\end{center}
\end{table}

The KPipe search for sterile neutrinos, which uses only the relative rate of events along the pipe, is helped by the fact that uncertainties associated with the absolute normalization of the event rate expectation are not relevant for this shape-only analysis.   This includes theoretical uncertainties in the kaon production and neutrino cross section.  Instead, the dominant uncertainty associated with the weight of each bin comes from the combined statistical uncertainty of the $\nu_{\mu}$ measurement and the CR background. In the sensitivity studies, we assume a CR background rate of 27~Hz over the entire detector. Further, there are two uncertainties associated with the neutrino baseline $L$: the creation point of the $\nu_{\mu}$ from the decaying $K^+$ has an uncertainty of 25~cm; the reconstructed position resolution, described in Section \ref{sec:recon}, has an uncertainty of 80~cm. There is no uncertainty associated with the energy reconstruction since the $\nu_\mu$ have a definite energy. We also include a total detection efficiency due to the selection cuts, dead-time, and escaping muons described in Section~\ref{sec:signal} of 77\%. A summary of the relevant experimental parameters and assumptions can be seen in Table~\ref{table:values}.

\begin{figure}[t!]
\centering
\begin{minipage}{1\linewidth}
  \includegraphics[width=\linewidth]{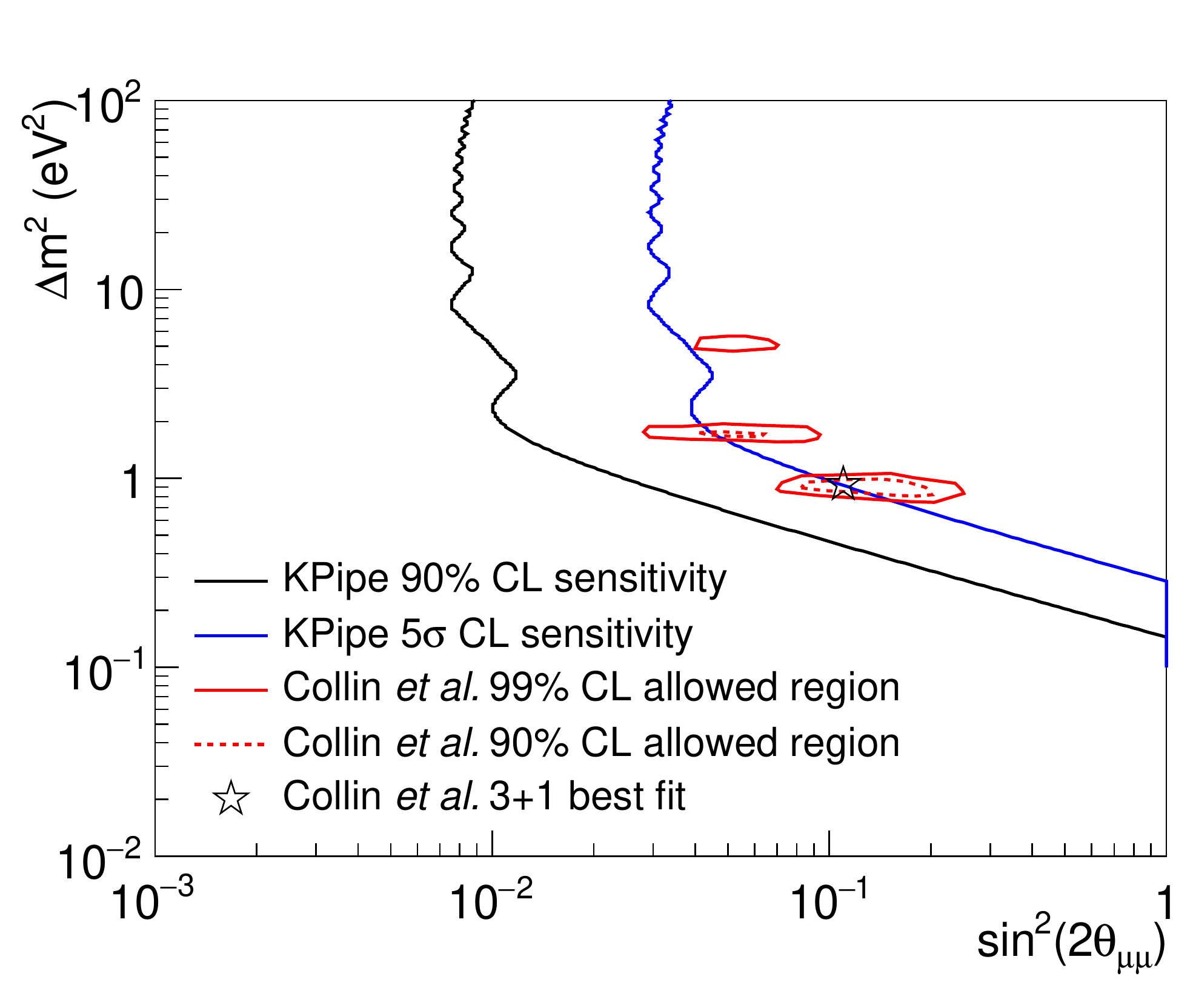}
\end{minipage}
\caption{The projected  sensitivity of KPipe to muon neutrino disappearance with 3 years of running, including the cosmic ray background, signal efficiencies, and reconstruction uncertainties described in the text. The red contours are the global allowed regions given by Collin \textit{et al.}~\cite{collin}.}
\label{fig:5s}
\end{figure}

Fig.~\ref{fig:5s} shows the projected $90\%$ and 5$\sigma$ sensitivity of KPipe to $\nu_{\mu}\rightarrow\nu_{\mu}$ for 3 years of running.  The global fit allowed regions, given in red, were produced using a new software package based on the previous work of Ignarra \textit{et al.}~\cite{sbl}. We refer to this work as ``Collin \textit{et al.}"~\cite{collin}. The fit includes the datasets described in Ref.~\cite{ignarra} with the exception of the atmospheric limit. The model parameters are explored using a Markov chain Monte-Carlo algorithm. Contours are drawn in a two-dimensional parameter space using 2 degree of freedom $\chi^2$ values for 90\% and 99\% probability. %The dashed contour is a combined raster scan exclusion at the 90\% confidence level of $\nu_\mu \rightarrow \nu_\mu$ disappearance experiments (CDHS, CCFR, MINOS, NOMAD, and MiniBooNE-SciBooNE combined). 
After 3 years of KPipe running, the 5$\sigma$ exclusion contour covers the best fit point at $\Delta m^2 = 0.93~\mathrm{eV}^2$ and $\sin^2(2\theta_{\mu \mu})= 0.11$. 

\begin{figure}[t!]
\centering
\begin{minipage}{1\linewidth}
  \includegraphics[width=\linewidth]{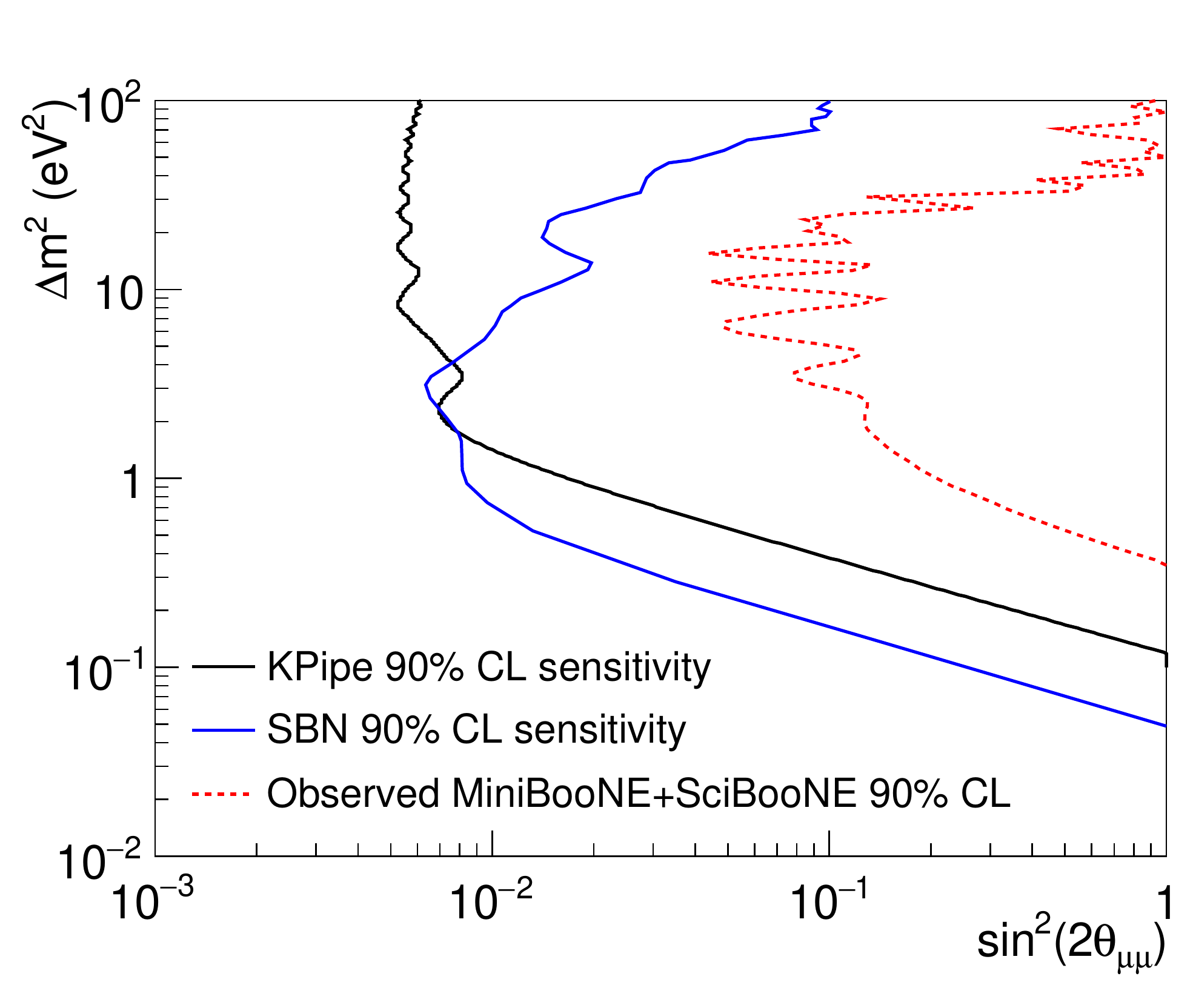}
\end{minipage}
\caption{The 90\% CL sensitivity of KPipe with 6~years of running, compared to the sensitivity from 6~years of the SBN program. The KPipe sensitivity estimate includes the cosmic ray background, signal efficiencies, and reconstruction uncertainties described in the text.}
\label{fig:90cl}
\end{figure}

Fig.~\ref{fig:90cl} shows a comparison between KPipe's predicted six~year $90\%$ sensitivity and the predicted sensitivity of SBN~\cite{fnal_sbn} assuming $6.6\times10^{20}$ POT (3 years) in SBND and the ICARUS-T600 and $13.2\times 10^{20}$ POT (6~years) in MicroBooNE. The dashed contour represents the combined $90\%$ excluded region based on the muon neutrino disappearance results of MiniBooNE and SciBooNE~\cite{MBnudisapp}. SBN and KPipe have similar sensitivity reach in the $\Delta m^2 = 1-4~\mathrm{eV}^2$ region, however SBN performs better at low-$\Delta m^2$ and KPipe at high-$\Delta m^2$; the complementarity between the experiments is clear.  

\section{Conclusion}

The J-PARC MLF facility provides a unique and intense source of neutrinos in the form of monoenergetic 236~MeV muon neutrinos coming from the decay-at-rest of positively charged kaons. The KPipe experiment seeks to take advantage of this source for a decisive $\nu_\mu$ disappearance search at high-$\Delta m^2$ in order to address the existing anomalies in this parameter space. The 120~m long, 3~m diameter liquid scintillator based active volume (684~ton) will feature 0.4\% photo-coverage for detecting these $\nu_\mu$ CC events in an attempt to discern an oscillation wave along the length of the detector.

In contrast to other neutrino sources, the KPipe neutrinos are dominantly monoenergetic. This provides a great advantage in searching for neutrino oscillations.  A neutrino (or antineutrino) induced double-coincidence muon signal detected with KPipe has a 98.5\% chance of being from a 236~MeV $\nu_\mu$ CC event. This simple fact allows the active detector requirements to be extremely modest, the systematic uncertainties to be practically eliminated, and the detector's energy resolution to be only a weak consideration. 

 %Since the neutrino energy is known, the event reconstruction only has to identify the existence of a double-coincidence KDAR neutrino interaction.  The detector and its capabilities can therefore be relatively modest and the systematic uncertainties are practically eliminated.

%This simple fact allows the active detector requirements to be extremely modest, as the detector's energy resolution -- typically critical -- is only a weak consideration. As a result, because the neutrino energy is known, the need to reconstruct details about the neutrino interaction lessens and so does the need for the usual experimental cost-drivers:  active detector components and read-out channels.

Within three years of running, KPipe will be able to cover the current global fit allowed region to 5$\sigma$. The sensitivity for a 6~year run at the J-PARC facility will enhance existing single experiment limits on $\nu_{\mu}$ disappearance by an order of magnitude in $\Delta m^2$. Such a measurement, when considered alone, or in combination with existing and proposed electron flavor disappearance and appearance measurements, can severely constrain models associated with oscillations involving one or more light sterile neutrinos.

\begin{center}
{ \textbf{Acknowledgments}}
\end{center}

The authors thank the National Science Foundation for support. This material is based upon work supported under Grant No. NSF-PHY-1205175.  TW also gratefully acknowledges the support provided by the Pappalardo Fellowship Program at MIT.  We thank Bill Louis, Marco Martini, and Stuart Mufson for discussions.

%Unused bibitems

%\bibitem{nova_ly}
%S. Mufson, ``Liquid Scintillator for NO$\nu$A".
%
%\bibitem{snoplus}
%M.C. Chen, 
%Nucl. Proc. Suppl., \textbf{145}, 65 (2005)
%Unused bibitems

%\bibitem{jaea}
%Sakamoto, Shinichi. Technical design report of spallation neutron source facility in J-PARC. Japan Atomic Energy Agency, Tokai, Ibaraki (Japan), 2012.
\end{document}